\documentclass[12pt,aps,showpacs,superscriptaddress,footinbib,preprint,noshowpacs]{revtex4-1}
\usepackage{color}
\usepackage{graphicx}
\usepackage{amsmath}
\usepackage{xcolor}
\usepackage{comment}
\usepackage{amssymb}
\usepackage{hyperref}
\usepackage[normalem]{ulem}

\usepackage{subfigure}
\usepackage{soul}

\begin{document}

\title{Bayesian unsupervised learning reveals hidden structure in concentrated electrolytes}

\begin{abstract}
Electrolytes play an important role in a plethora of applications ranging from energy storage to biomaterials. Notwithstanding this, the structure of concentrated electrolytes remains enigmatic. Many theoretical approaches attempt to model the concentrated electrolytes by introducing the idea of ion pairs, with ions either being tightly `paired' with a counter-ion, or `free' to screen charge. In this study we reframe the problem into the language of computational statistics, and test the null hypothesis that all ions share the same local environment. Applying the framework to molecular dynamics simulations, we show that this null hypothesis is not supported by data. Our statistical technique suggests the presence of distinct local ionic environments; surprisingly, these differences arise in like charge correlations rather than unlike charge attraction. 
The resulting fraction of particles in non-aggregated environments shows a universal scaling behaviour across different background dielectric constants and ionic concentrations.
\end{abstract}

\author{Penelope Jones}
\email{pj321@cam.ac.uk}
\affiliation{Department of Physics, University of Cambridge, CB3 0HE, Cambridge, United Kingdom}

\author{Fabian Coupette}
\affiliation{Institute of Physics, University of Freiburg, Hermann-Herder-Stra\ss{}e 3, 79104 Freiburg im Breisgau, Germany}

\author{Andreas H\"{a}rtel}
\email{andreas.haertel@physik.uni-freiburg.de}
\affiliation{Institute of Physics, University of Freiburg, Hermann-Herder-Stra\ss{}e 3, 79104 Freiburg im Breisgau, Germany}

\author{Alpha A. Lee}
\email{aal44@cam.ac.uk}
\affiliation{Department of Physics, University of Cambridge, CB3 0HE, Cambridge, United Kingdom}

\maketitle

Understanding the correlations between charged objects in electrolytes is important to applications such as colloid science \cite{hansen_effective_2000, barrat_basic_2003} and energy storage in supercapacitors \cite{simon_materials_2008, simon_perspectives_2020}. The structure of dilute electrolytes is well-understood: the mean field Debye-H\"{u}ckel theory, derived almost a century ago, accurately predicts a decrease in electrostatic screening length as ion concentration increases \cite{huckel_zur_1923}. However, the physical picture is less clear for concentrated electrolytes. Theoretical studies have systematically analysed the structure of electrolytes beyond mean-field theory \cite{attard_asymptotic_1993, carvalho_decay_1994, doi:10.1063/1.469166}. These pioneering works predict an increase in screening length after reaching a minimum when the screening length is of the order of the ion diameter in concentrated electrolytes. 

Recently, a series of experimental studies using the Surface Force Balance (SFB) apparatus revealed that the screening length is longer than theoretically predicted, reaching up to 10 nm (an order of magnitude greater than the ion diameter) in solvent-free ionic liquids \cite{gebbie_ionic_2013, gebbie_long-range_2015, smith_electrostatic_2016, gaddam_electrostatic_2019}. The screening length follows a simple scaling behaviour that is linear in ion concentration and Bjerrum length, and is universal across organic and inorganic electrolytes \cite{lee_scaling_2017}. 

The physical mechanism that triggers this surprisingly long screening length is still elusive. Atomistic and coarse grained simulations have observed a screening length that seems to agree with the short-range `structural force' observed in experiments, but not the long decay length \cite{coles_correlation_2020}. Many theoretical approaches attempt to explain the data by introducing the idea of ion pairs \cite{gebbie_ionic_2013, gebbie_long-range_2015} or effective dielectric constants \cite{kjellander_multiple_2020}. The argument is that most cations and anions are bundled up as neutral pairs in concentrated electrolytes, thus the electrolyte comprises a low concentration of `free' ions solvated in a liquid of effectively neutral `paired' ions. However, ion pairs are usually defined based on a cutoff distance selected to fit the data. It is unclear why ions closer than a certain distance are physically distinct from the rest, and how this affects the dielectric constant. 

In this paper, we revisit the physical picture of ion pairing using Bayesian inference and unsupervised machine learning. Rather than attempting to predict and explain the anomalously long screening length, our aim is to understand the concept of ion pairing. Our conjecture is that if there were ion pairs, one would expect the existence of two distinct local ionic environments, one corresponding to ions in a `paired' state and another corresponding to ions in a `free' state. The question then becomes whether the hypothesis that all ions share one environment is more probable than the hypothesis that there are multiple ionic environments. As such, we reframe the ion-pair hypothesis as Bayesian hypothesis testing \cite{jeffreys_theory_1961}, sidestepping any phenomenological distance-based cutoff. 

We will first discuss the representation of local ionic environments using the local pair distribution function and introduce the Bayesian Gaussian mixture model as a tool to determine the most probable number of statistically distinct local environments. We will then analyse molecular dynamics simulation results of the solvent primitive model, used as a surrogate model of the bulk electrolyte. At intermediate concentrations, we infer the presence of multiple statistically distinct ionic environments, whose differences are grounded in the form of like charge correlations rather than unlike charge pairing. We conclude by mapping the inferred fraction of particles in non-aggregated environments to an effective screening length, and by discussing the resulting collapse of our data onto one single curve.

\section{Representation and clustering of local environments}

\subsection{Description of local ionic environment}
We describe the local environment of some central ion $a$ of type $A$ by its correlations with surrounding ions of type $B$
\begin{equation}
    \tilde{g}_{aB}(r) = \frac{V}{N_B} \sum_{b = 1}^{N_B} \mathcal{N}(r;  |\mathbf{r}_{ab}|, \sigma^2),
    \label{eqn:descriptor}
\end{equation} 
where $V$ is the system volume, $N_B$ is the number of type $B$ ions, and each neighbouring ion $b$ is represented by a normal distribution $\mathcal{N}$, with mean and standard deviation equal to the inter-ion distance $|\mathbf{r}_{ab}|$ and the ion radius $\sigma$. Our descriptor, $\tilde{g}_{aB}(r)$, is related to the pair distribution function (PDF) 
\begin{equation} \label{eq:unsmoothed_pdf}
    g_{AB}(r)=\frac{1}{N_A}\sum_{a=1}^{N_A}\frac{V}{N_B}\sum_{b=1}^{N_B}\left<\delta(r - |\mathbf{r}_{ab}|)\right>,
\end{equation} 
where $<\cdot>$ denotes ensemble average \cite{frenkel_understanding_1997}, except we smooth the delta function to a Gaussian, and remove the ensemble average as we are considering the local environment at a particular snapshot, around a particular ion. 

The representation is vectorised by discretising $r$ into $D$ bins; the resultant descriptor for each central ion $\mathbf{x}$ is formed from the concatenation of representations of both like and unlike charge correlations and is thus of dimensionality $d = 2D$. From $F$ frames of molecular dynamics simulations, each with some number of type $A$ ions $N_{A}$ in the simulation box, we collect a total of $N = N_{A}F$ local ionic environments, $\{\mathbf{x}_i\}_{i=1}^{N}$. The simulation methodology and numerical details are provided in Appendix \ref{simulations}.

\subsection{Pattern identification with Bayesian unsupervised learning}
Having collected the local ionic environments to form the dataset $\mathbf{X} = \{\mathbf{x}_n\}_{n=1}^{N}$, we next consider the number of statistically significant groups into which the $N$ datapoints fall. If statistically speaking, ions do not all inhabit the same local environment, we can `group' the inhabited environments such that differences within each group are much smaller than differences between groups.

This grouping problem can be solved using Bayesian inference \cite{mackay_information_2003, bishop_pattern_2006}. We first posit that each data point $\mathbf{x}_n$ is generated by the sum of $K$ independent probability distributions, each describing a local environment. We then infer both the most probable number of distributions required to adequately describe the data, $K^*$, and the most probable parameters of those distributions. $K^*$ is physically interpreted as the number of statistically distinct environments in the system, a key parameter that could confirm or reject the ion pair picture. 

For ease of interpretation, we consider a weighted sum of $K$ Gaussian distributions,
\begin{equation}
p(\mathbf{x}_n|K, \boldsymbol{\Theta}_K) = \sum_{k=1}^{K} \pi_{k} \mathcal{N}(\mathbf{x}_n;  \boldsymbol{\mu}_{k}, \boldsymbol{\Lambda}^{-1}_{k}) \label{eqn_gmm}
\end{equation}
where $\boldsymbol{\Theta}_K = \{\pi_k, \boldsymbol{\mu}_{k}, \boldsymbol{\Lambda}_{k}\}_{k=1}^K$ are the parameters of the $K$-environment model; $\pi_{k}$ defines the fraction of ions inhabiting the $k$th local environment, and $\boldsymbol{\mu}_{k}$ and $\boldsymbol{\Lambda}_{k}$ define the mean of the $k$th local environment (physically, the local ionic environment of the typical ion in that group) and the precision of environments within that group respectively. The parameters $\boldsymbol{\Theta}_K$ are considered random variables; an approximation will be found to the posterior distribution $p(\boldsymbol{\Theta}_K|\textbf{X}, K)$, which is the probability of the parameters given the observed data. The form of this Bayesian Gaussian mixture model is given in Appendix \ref{BGMM}.

The optimal number of distributions, $K^*$, is that which maximises the probability $p(\mathbf{X}|K)$ of the observed data given a $K$-environment model. $p(\mathbf{X}|K)$ is the \textit{marginal likelihood}; it quantifies how well a model can explain the observed data and is of fundamental importance in Bayesian model selection \cite{kass_bayes_1995}. 

The Bayesian paradigm is that the quality of a $K$-environment model is not measured by how well the `best' parameters within a model class fit the data, but whether the model, averaged over all possible model parameters, can effectively fit the data. This is seen in the explicit marginalisation over all model parameters, $\boldsymbol{\Theta}_K$:
\begin{align}
p(\mathbf{X}|K) = \int p(\mathbf{X}| \boldsymbol{\Theta}_K, K) p(\boldsymbol{\Theta}_K | K) d\boldsymbol{\Theta}_K. \label{eqn20}
\end{align}
Crucially, the marginal likelihood incorporates the Occam's razor effect \cite{mackay_bayesian_1991}, implicitly penalising overly complex models. It will be maximised by the simplest model that can adequately explain the data.

We shall use the Bayes factor
\begin{equation}
BF(K'|K) = \frac{p(\mathbf{X}| K')}{p(\mathbf{X}| K)},
\end{equation}
whose calculation is central in Bayesian hypothesis testing \cite{jeffreys_theory_1961, kass_bayes_1995}, to obtain a quantitative measure of the evidence in favour of a model with $K'$ environments as against another with $K$ environments. Rather than computing the integral (\ref{eqn20}), which is numerically intractable, in this paper we employ variational inference \cite{jordan_introduction_1998, wainwright_graphical_2007, blei_variational_2017} to compute an approximation to the marginal likelihood and corresponding Bayes factor, inspired by the statistical physics of mean field theory \cite{mackay_information_2003}. A summary of this approach is provided in Appendix \ref{VI}. The code used can be found at \url{https://github.com/PenelopeJones/electrolytes}.

\section{Results and Discussion}
\subsection{Validation on toy systems}\label{validation}

\begin{figure}[h]
    \centering
    \includegraphics[width=0.6\linewidth]{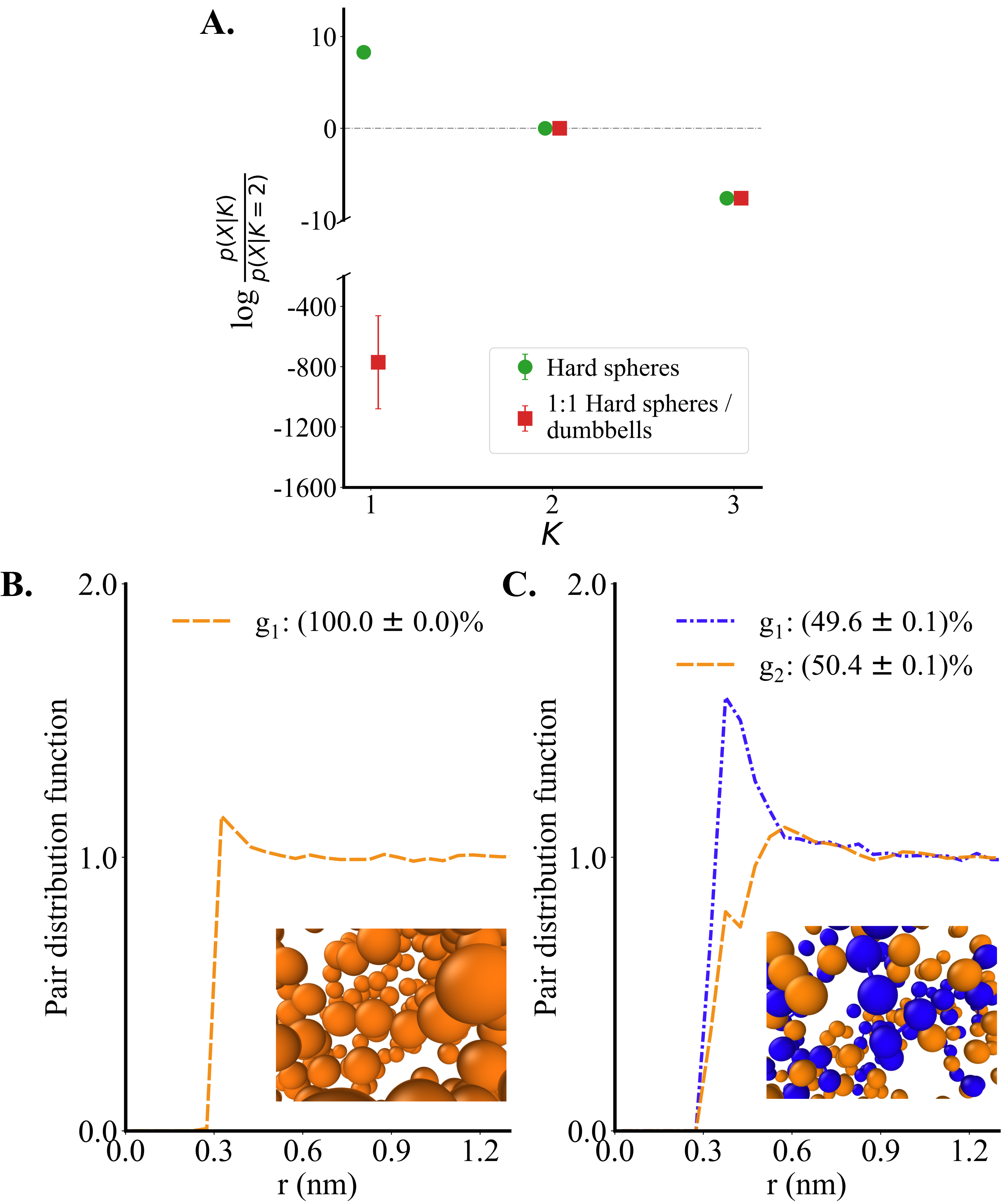}
    \caption{The method is validated on two systems: one comprising only hard spheres, and one comprising hard spheres and dumbbells in a 1:1 ratio. \textbf{A.} The number of statistically distinct environments, $K^*$, is that which maximises the marginal likelihood $p(\mathbf{X}|K)$; $K^*=1$ and $2$ for the first and second system respectively. \textbf{B/C.} Unsmoothed PDFs for the identified environment(s), for \textbf{B.} the hard spheres system, and \textbf{C.} the 1:1 hard spheres / dumbbells system.}
    \label{fig:known}
\end{figure}

Before turning to analyse electrolytic solutions, we first illustrate how our methodology can be applied to systems where the answer is well-known. Through this set of toy problems, we validate the choice of hyperparameters in the Bayesian prior which we will then apply to understand the structure of electrolytes.  Technical details are provided in Appendix \ref{BGMM}.

We consider two simple systems: The first is a system of 100\% neutral hard spheres and the second comprises unbonded hard spheres and bonded hard spheres in a $1:1$ ratio; both systems have the same total concentration of hard spheres, although in the latter case half of the hard spheres are connected by a harmonic spring. The question is: if we assume we do not know which particles are bonded and which particles are free, can the algorithm decipher that there are two distinct environments, and correctly deduce the proportion of particles in each environment? Figure \ref{fig:known} shows that in agreement with intuition, the most probable number of statistically distinct local environments, $K^*$, is inferred to be one for the system comprising only neutral hard spheres, but two for the system comprising neutral hard spheres and dumbbells in a $1:1$ ratio. 

We then study the differences in these statistically distinct environments. Rather than studying the posterior over the model parameters themselves, it is more intuitive to look at the widely studied unsmoothed PDF, following \eqref{eq:unsmoothed_pdf}, averaged over all particles classified as inhabiting each environment. These are shown for the two systems in Figure \ref{fig:known}. For the first system, the recovered mean ($g_1$) corresponds exactly to the `true' mean environment for the system. For the second system, the recovered environments are seen to correspond to the `bonded' ($g_1$) and `unbonded' ($g_2$) environments, with the correct proportion of ions classified as each. It should be noted that the recovered mean PDFs do not exactly correspond to those of the dumbbells and hard spheres, but the majority of dumbbells are classified as `bonded' and the majority of hard spheres are classified as `unbonded'.

\begin{figure*}[t]
\centering
\includegraphics[width=0.8\linewidth]{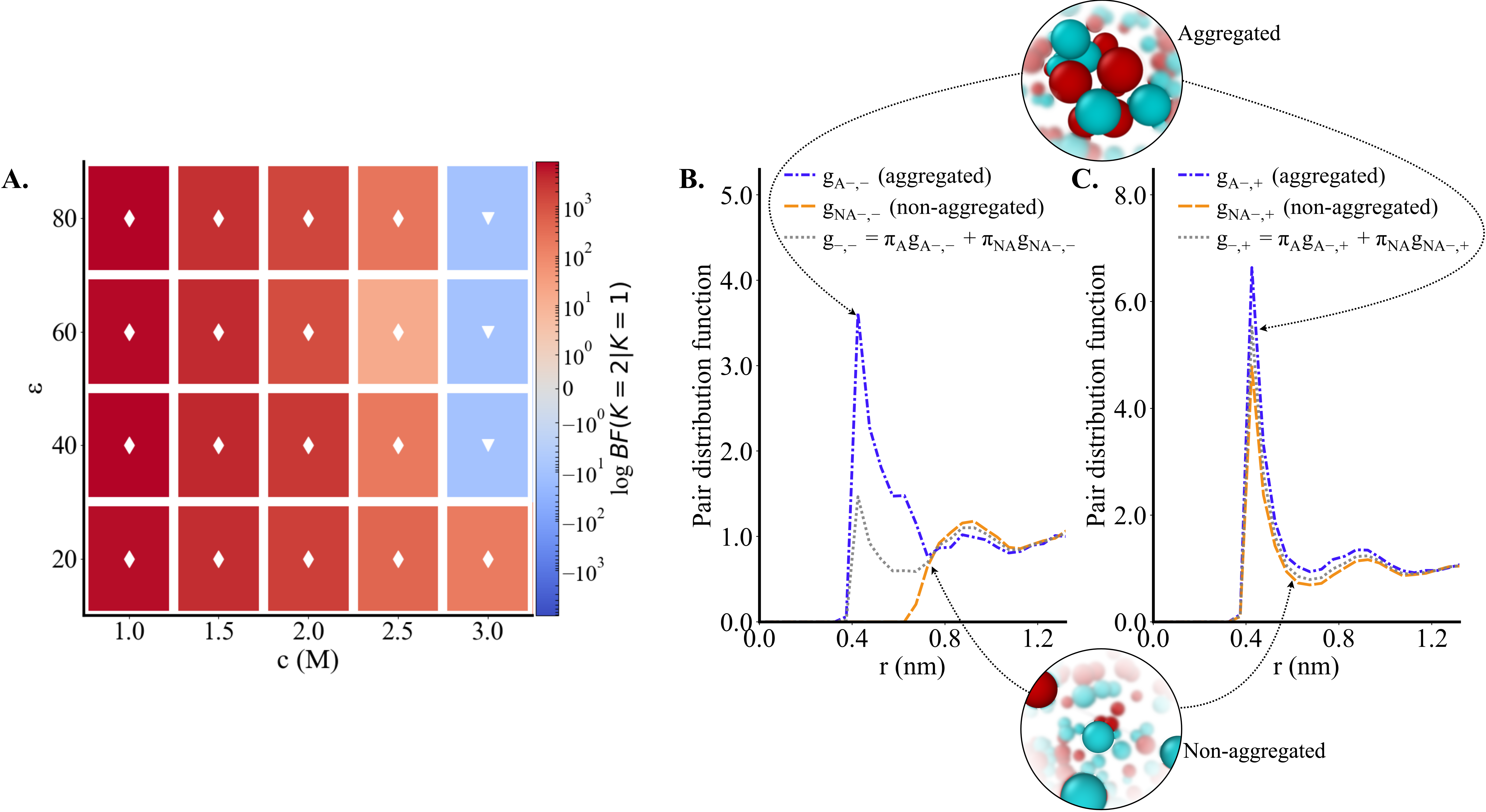}
\caption{Analysis of the Solvent Primitive Model. \textbf{A.} At intermediate concentrations, the number of statistically distinct environments, $K^*$, is 2; at higher concentrations $K^* = 1$. Here, the stronger the evidence supporting a two-environment model (as against a one-environment model), the redder the shading ($\blacklozenge$: $BF(2|1)>0$); conversely the stronger the evidence supporting a one-environment model, the bluer the shading ($\blacktriangledown$: $BF(2|1)<0$). \textbf{B/C.} The two distinct environments are distinguished as \textit{aggregated} and \textit{non-aggregated} in real space. Here, the unsmoothed PDFs, averaged over all ions classified as inhabiting each environment are shown for the $c$ = 1.0 M, $\epsilon$=80 system. \textbf{B.} and \textbf{C.} show like and unlike charge correlations respectively. Differences in environments originate in like charge correlations.}
\label{fig:electrolyte}
\end{figure*}

\subsection{Monovalent electrolytic solutions}
\begin{figure}[!ht]
    \centering
    \includegraphics[width=0.4\linewidth]{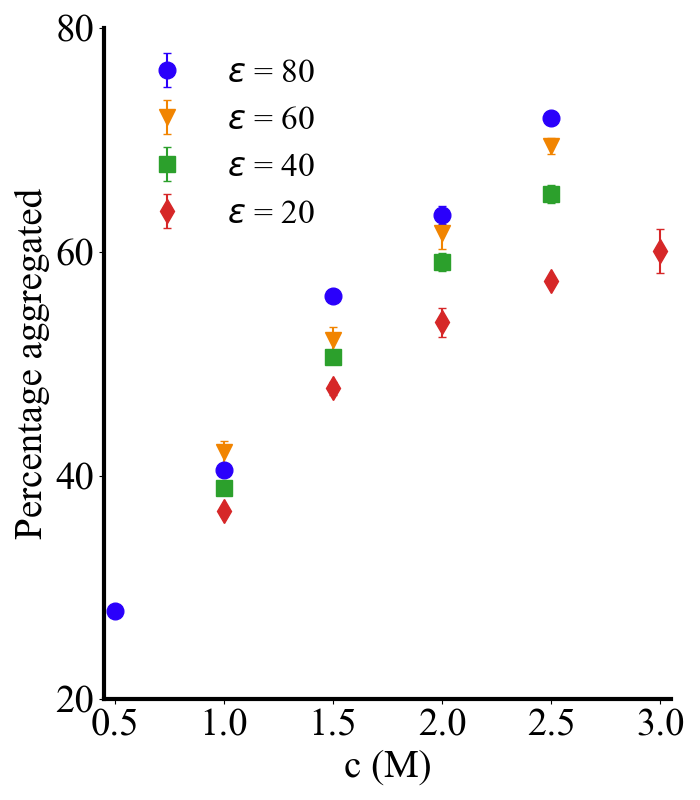}
    \caption{In the Solvent Primitive Model, the proportion of ions classified as `aggregated' increases with ionic concentration $c$ and dielectric constant $\epsilon$.}
    \label{fig:n_variation}
\end{figure}

Having corroborated the ability of the model to infer the correct number of environments in simple physical systems, we now apply the same technique to probe the structure of bulk monovalent electrolytes, modelled using the Solvent Primitive Model (SPM) as outlined in Appendix \ref{simulations}. Figure \ref{fig:electrolyte} summarises our key findings: At intermediate concentrations the evidence strongly supports a two environment model as against a single environment model. At higher concentrations these two statistically distinct environments become indistinguishable, and the evidence supports a single environment model. We note that in experiments, the screening length as a function of ion concentration appears to deviate from the scaling behaviour at concentrations near saturation/neat ionic liquids \cite{smith2017switching}. 

Scrutiny of the two distinct environments inferred at intermediate concentrations reveals that surprisingly, the preeminent difference between environments arises in the like charge correlations. This contrasts with the differences in unlike charge correlations implied by the ion-pair hypothesis. The two environments are physically distinguished as being `aggregated' and `non-aggregated'. In the first case, several charges are bundled together in close proximity; in the second case there will be \textit{at most} another unlike charge in the local vicinity. 

Interestingly, for all concentrations $c$ and dielectric constants $\epsilon$ for which a two environment model is found to be most probable, the same qualitative differences in environment are identified, but the proportion of ions classified as aggregated increases with both $c$ and $\epsilon$, as shown in Figure \ref{fig:n_variation}. This relationship is rationalised by the reduced electrostatic repulsion between like charges at larger $\epsilon$, and by the reduced ability to maintain large distances between like charges at larger $c$.

\begin{figure}[t]
    \centering
    \includegraphics[width=0.4\linewidth]{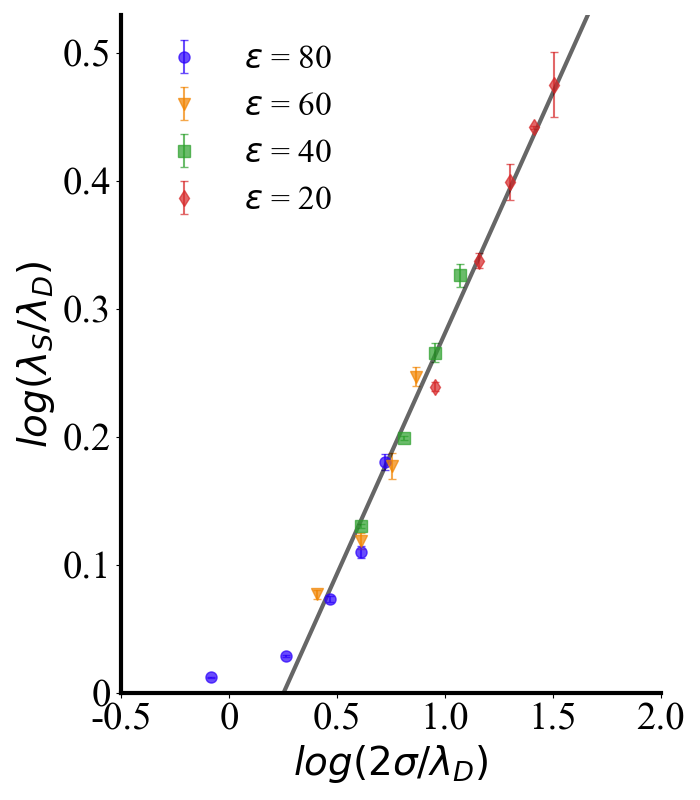}
    \caption{The effective screening length, computed using the fraction of aggregated ions, displays a scaling behaviour that holds across different background dielectric constants and ion concentrations.}
    \label{fig:screening_length_scaling}
\end{figure}

We next consider the relationship between aggregated ions and the screening length. In the classical ion pair model, ion pairs are considered neutral species that effectively reduce the ion concentration in the solution, thus increasing the screening length according to Debye-H\"{u}ckel theory. If we were to consider aggregated ions playing the role of ion pairs, one might posit that the effective screening length takes the form, analogous to the Debye length, 
\begin{equation}
\lambda_S = \sqrt{\frac{\epsilon_{\mathrm{eff}}\epsilon_0 k_B T}{2e^2 c_{\mathrm{eff}}}}  
\end{equation}
where $\epsilon_{\mathrm{eff}}$ is the effective dielectric constant of the mixture of non-aggregated ions and ion aggregates, and $c_{\mathrm{eff}}$ is the concentration of non-aggregated ions.
We model the solution as a dielectric continuum in which ion aggregates, with dielectric constant $\epsilon_A$ and molecular fraction $\phi$, exist in a dielectric background with dielectric constant $\epsilon$. Note that for each system, $\phi$ is determined directly from the approximate posterior over model mixing coefficients $\{\pi_k\}_{k=1}^K$ (c.f. Figure \ref{fig:n_variation}). The effective concentration of non-aggregated ions is then
\begin{equation}
    c_{\mathrm{eff}} = c(1 - \phi).
\end{equation}
We compute $\epsilon_{\mathrm{eff}}$ from the Bruggeman equation for dielectric mixing \cite{markel2016introduction}:
\begin{equation}
\epsilon_{\mathrm{eff}} = \frac{b + \sqrt{8 \epsilon_A \epsilon + b^2}}{4}
\end{equation} 
where $b= (3 \phi-1) \epsilon_A + (2-3 \phi)\epsilon$. Following the scaling analysis of the experimental data \cite{lee_scaling_2017}, we analyse the data by plotting $\lambda_S$/$\lambda_D$ against $2\sigma/\lambda_D$, with $\lambda_D$ the Debye length computed using $\epsilon$ and $c$, and $\sigma$ the ion radius. The only unknown constant is $\epsilon_A$, the dielectric constant of the ion aggregates. We fit $\epsilon_A$ to maximise the extent to which the data collapses. Note that this is a single parameter fit, collapsing data from simulations at 4 different dielectric constants, with multiple ion concentrations at each dielectric constant. Figure \ref{fig:screening_length_scaling} shows that the data collapses satisfactorily for $\epsilon_A \sim 21.1$, in qualitative agreement with experimental data \cite{lee_scaling_2017}. However, the scaling exponent is $0.376 \pm 0.011$, which is different to the cubic scaling observed in experiments as well as the scaling observed in simulations \cite{coles_correlation_2020}. 

\section{Conclusion}
In summary, we used Bayesian unsupervised learning to infer the number of statistically distinct local ionic environments in bulk monovalent electrolytes. Surprisingly, our results indicate the presence of multiple statistically distinct environments at intermediate concentrations, whose differences originate in like charge correlations as opposed to unlike charge correlations which would be implied by the ion-pair hypothesis. 

The presence of a scaling relationship suggests that there might be underlying general physical insights which could be inferred from concepts such as aggregated ions and effective ion concentration, revealed by the unsupervised learning approach. The effective dielectric constant of the ion aggregate that collapses a wide range of simulation data is inferred to be $\epsilon_{A} \approx 21$. This is perhaps unsurprising as it is in the range expected for room temperature ionic liquids, which are pure ionic melts. Nonetheless, we note that the exponent of the scaling relationship obtained from our unsupervised learning analysis is different from what is observed experimentally. 

Perhaps most importantly, our results suggest that meaningful physical insights can be elucidated in soft matter systems by studying the statistical \textit{differences} between local environments in the same macroscopic system, as opposed to more widely used statistical \textit{averages}. We suggest this could be a novel way to recharacterise previously hidden order in soft matter and ionic systems. 

\begin{acknowledgments}
P.J. and A.A.L. acknowledge the support of the Winton Programme for the Physics of Sustainability. P.J. acknowledges the support of the Ernest Oppenheimer Fund. A.H. and F.C. acknowledge funding by the Deutsche Forschungsgemeinschaft (DFG, German Research Foundation) - project numbers 406121234; 404913146, respectively. 
The software package Ovito was used for the visualisation of environments in real space \cite{stukowski_visualization_2009}.
This work was performed using resources provided by the Cambridge Service for Data Driven Discovery (CSD3) operated by the University of Cambridge Research Computing Service, provided by Dell EMC and Intel using Tier-2 funding from the Engineering and Physical Sciences Research Council (capital grant EP/P020259/1), and DiRAC funding from the Science and Technology Facilities Council (www.dirac.ac.uk).
\end{acknowledgments}

\section*{Data Availability Statement}
The data that support the findings of this study are openly available in 10.5281/zenodo.4015136 at
https://zenodo.org/deposit/4015136. 

\appendix

\section{Data production and feature extraction} 
\label{simulations}
Using the ESPResSO package \cite{weik_espresso_2019}, molecular dynamics simulations were carried out both for the `toy' systems and for the Solvent Primitive Model (SPM) \cite{grimson_forces_1982}. For both toy systems, molecules were modelled as neutral hard spheres of radius $\sigma$ = 0.15 nm. Hard particle interactions were modelled by a Weeks-Chandler-Andersen (WCA) potential \cite{andersen_relationship_1971, weeks_role_1971}, a purely repulsive truncated and shifted Lennard-Jones potential with prefactor $10^4 \ k_{\textrm{B}}T$ and cut-off distance at $2\sigma$, where $k_B$ is the Boltzmann constant and $T$ is the temperature. In the second toy system, `dumbbells' were formed by connecting 50\% of the spheres to another using additional harmonic bonds of equilibrium separation 0.4 nm and prefactor 15 $k_{\textrm{B}}T$/nm$^{\textrm{2}}$. Smoothed PDFs were calculated for distances [0.15 nm, 1.2 nm] using a bin size of 0.15 nm and normalised with linear scaling, such that the resulting data had zero mean and range [-1, 1].

To model the electrolytic system, we use the SPM, which is one of the simplest molecular solvent models, and has been used extensively to study interactions within electrolytic solutions at surfaces \cite{tang_threecomponent_1992, zhang_simulations_1993, boda_capacitance_2000} where packing effects matter. More recently, it has been applied to investigate the underscreening effect in the bulk of ionic liquids and concentrated electrolytes \cite{rotenberg_underscreening_2018, coupette_screening_2018} and to explain an experimentally observed switch in the decay behaviour of correlations \cite{coupette_screening_2018}. 

In the SPM, cations, anions and solvent molecules were modelled as hard spheres of radius $\sigma$ = 0.2 nm with electric charges $+e, -e$ and $0$ respectively. Simulations were performed at constant temperature $T = 300$ K and a range of concentrations $c$ and relative permittivities $\epsilon$, i.e.  $(c, \epsilon) \in$ [0.5 M, 3.0 M] $\times$ \{20, 40, 60, 80\}. Hard particle interactions were again modelled by a WCA potential, and electrostatic interactions were treated by the P3M method \cite{weik_espresso_2019, hockney_computer_1988}. The smoothed PDFs $g_{+-}$ and $g_{--}$ were calculated for distances [0.2 nm, 1.0 nm] using a bin size of 0.2 nm and normalised as described above. Only the anionic environments are studied due to charge reversal symmetry.

\section{The Bayesian Gaussian mixture model}
\label{BGMM}
In the Gaussian mixture model, it is assumed that the observed dataset $\mathbf{X} = \{\mathbf{x}_1, ..., \mathbf{x}_N \}$ is drawn from an underlying distribution comprised of a linear superposition of $K$ independent Gaussian distributions, each of which has an associated mean $\boldsymbol{\mu}_k$, precision $\boldsymbol{\Lambda}_k$ (note that the precision is the inverse of the covariance $\boldsymbol{\Sigma}_k$), mixing coefficient $\pi_k$, where $\pi_k$ represents the probability of an observed data point being drawn from the $k$th distribution. Each observation $\mathbf{x}_n$ has an associated latent variable $\mathbf{z}_n$ of dimensionality $K$, where $z_{nk}$ is the label describing the mapping of datapoint $\mathbf{x}_n$ to cluster $k$. Thus the latent variables can be denoted by $\mathbf{Z} = \{\mathbf{z}_1, ..., \mathbf{z}_N\}$. The distribution over $\mathbf{Z}$ given $\boldsymbol{\pi} = \{\pi_k\}_{k=1}^K$ is
\begin{align}
    p(\mathbf{Z}|\boldsymbol{\pi}) = \prod_{n = 1}^N \prod_{k = 1}^K \pi_k^{z_{nk}}
\end{align}
and the distribution over $\mathbf{X}$ given the latent variables and model parameters is
\begin{align}
    p(\mathbf{X}|\mathbf{Z}, \boldsymbol{\mu}, \boldsymbol{\Lambda}) = \prod_{n = 1}^N \prod_{k = 1}^K \mathcal{N}(\mathbf{x}_n|\boldsymbol{\mu}_k, \boldsymbol{\Lambda}^{-1}_k)^{z_{nk}}.
\end{align}
In the Bayesian Gaussian mixture model, it is further assumed that the parameters $\boldsymbol{\Theta}_K = \{\pi_k, \boldsymbol{\mu}_{k}, \boldsymbol{\Lambda}_{k}\}_{k=1}^K$ are themselves random, with distributions parameterised by hyperparameters to be determined. The aim is to find the posterior distribution $p(\boldsymbol{\Theta}_K|\textbf{X}, K)$ of these model parameters given $\mathbf{X}$. 

To compute the posterior, it is necessary to specify a prior distribution over $\boldsymbol{\Theta}_K$. We use the prior 
\begin{equation} p(\boldsymbol{\pi}) = \textrm{Dir}(\boldsymbol{\pi}|\boldsymbol{\alpha}_0) \label{eqn_prior1}
\end{equation}
\begin{equation} p(\boldsymbol{\mu| \Lambda})p(\boldsymbol{\Lambda}) = \prod_{k = 1}^K \mathcal{N}(\boldsymbol{\mu}_k|\boldsymbol{m_0,  (\Lambda_k}\beta_0)^{-1})\mathcal{W}(\boldsymbol{\Lambda_k}|\mathbf{W_0}, \nu_0). \label{eqn_prior2}
\end{equation}
Using this model, the only manual selection is that of the hyperparameters $\{\alpha_0, \boldsymbol{m_0}, \beta_0, \mathbf{W_0}, \nu_0\}$. We select uninformative hyperparameters (those that lead to a broad distribution over model parameters): $\alpha_0 = 1.0$, $\beta_0 = 1.0\times 10^{-11}$, $\mathbf{W_0} = \mathbb{I}$, $\boldsymbol{m_0} = \mathbf{0}$ and $\nu_0 = d$, where $d$ is the dimensionality of $\mathbf{x}$. We check for coherence of the resulting prior distribution via validation on well understood physical systems as described in the main text.

\section{Variational inference} 
\label{VI}
Calculation of the marginal likelihood requires integration over all parameters of the model which is intractable for most models of interest. This problem can be circumvented using variational inference \cite{blei_variational_2017}, whereby the variational distribution $q_{\phi}(\boldsymbol{\Theta}_K)$ is introduced and used to obtain a lower bound to the marginal likelihood, termed the evidence lower bound (ELBO):
\begin{equation}
    \mathcal{L}(\phi) = \ \int q_{\phi}(\boldsymbol{\Theta}_K) \log\left(\frac{p(\mathbf{X}, \boldsymbol{\Theta}_K|M_K)}{q_{\phi}(\boldsymbol{\Theta}_K)}\right) d\boldsymbol{\Theta}_K.
\end{equation}
Maximisation of the ELBO with respect to variational parameters $\phi$ jointly obtains an approximation to the marginal likelihood and the posterior of the model parameters $p(\boldsymbol{\Theta}_k|\mathbf{X}, M_k)$. Indeed, when $q_{\phi}(\boldsymbol{\Theta}_k) = p(\boldsymbol{\Theta}_k|\mathbf{X}, M_k)$ the lower bound is exact.

% Bibliography
%\bibliography{references.bib}

\end{document}